# Numerical Simulation of Thermal Energy Storage using Phase Change Material

Abhishek Rai, N.S Thakur, Deepak Sharma

Department of Mechanical Engineering, NIT Hamirpur, H.P.-177005, India

**Highlights:**

- CFD modelling and simulation of Thermal Energy Storage using Phase Change Material.
- Gallium is used as Phase Change Material due to its high thermal conductivity than paraffin.
- The design with fins gives higher heat transfer rate with optimized number of heat sources.

**Abstract:**

In recent years, the researchers focused on the storage of direct sunlight in the form of heat by utilizing the temperature change with various techniques. The principal task in Thermal Energy Storage is the time span of charging and discharging. The time span to store the heat without losing temperature is additionally significant. Bearing the various innovations, thermal storages can store energy for an appreciable period of time to balance the demand by giving the same amount of heat as stored with very little loss in form of heat convection. This study includes the design optimization of Thermal Energy Storage (TES) in the form of the cylindrical cavity with the use of Gallium as a Phase Change Material (PCM). The process involves the use of CFD simulation and the design of five different models on ANSYS Fluent as a software tool. To optimize the design of each model with different geometries, they have been examined under similar operating conditions. The first model has no fins and comprises of four heat sources, while the other four model consists of fins on the heat source surface with the variation in the number of heat sources and design of fins. The complete study has been carried out at 343K as the heat source temperature which is constant for the fin's surface as well. The external surface of the cylinder is assumed to be adiabatic in nature with an atmospheric temperature of 298K. The operating and the boundary conditions are kept to be the same for the analysis of each model that has been examined at different time period. The results show that the design with fins gives better results than the design with no fins due to increased heat transfer surface area. The comparison of models for better optimization shows that model 4 with three heat sources gives almost similar results as given by Model 2 with four heat sources. In further study, we examine the Model 5 and conclude that the results are not useful as the time of charging of PCM increases. The study clearly shows that the design of the Model 4 with three heat sources gives the optimized results.

**Keywords:** Phase Change Materials (PCM), Thermal Energy Storage (TES), CFD, Solar energy, Heat source.

# 1. Introduction

Solar energy is one of the most abundant source of energy on the earth. Free availability of solar energy on various parts of the earth is its main benefit. Transformation of solar energy can be utilized for thousands of years by people traditionally for different purposes like heating, cooking, drying. The solar energy can be utilized directly for numerous purposes and is served as most popular source of renewable energy. Nowadays solar energy treated as the perfect replacement of energy derived by fossil fuels. Sustainable energy source has vitality that they can be gathered from inexhaustible resources. They are normally renewed on human timescale such as wind, sunlight, rain, tides, waves, geothermal etc. One of the major possible sector of using renewable energy is in the form of solar energy. Solar energy can be considered as the best replacement of electrical forms of energy. The main focus of using solar energy is its utilization with best techniques possible to reduce loss of energy in the form of heat. As we all know energy can neither be created nor be destroyed, can only be converted from one form to another. In the same manner the solar energy can be converted from one form to any other form of energy. Due to the presence of photons in solar radiation the solar energy can be basically used to convert in the form of electricity by the use of solar cells, but the solar energy can also be directly absorbed by some means. Solar radiations after passes the earth's atmosphere available in the form of visible radiation and infrared radiation. Excess of solar energy falls on earth's surface can be absorbed by the water in seas and oceans in the form of thermal heat. The various researchers all over the globe studied about the thermal energy storage (TES) and its applications with the use of other materials that can absorb energy. Generally, the studies include the applications of TES with phase change materials (PCM). The design of TES with PCM is the most important part. T. Bouhal et al. [1] (2018) in this paper enhancement of thermal energy storage (TES) using gallium as PCM in a cylindrical cavity with heating source was simulated by CFD. The focus is to optimize the geometry for the given temperature of heat source. To compare four fins were used on heat source under same operating conditions which improves the melting time. The assumption includes the negligible convective effect and thermal hysteresis. The temperature is higher than the melting point of PCM, effect of heat source with different geometries investigated using solidification and melting model. Ammar M. Abdulateef et al. [2] (2018) presented the study of PCM experimentally and numerically in a triplex tube heat exchanger with fins. Two types of extended surfaces used longitudinal and triangular. Shell and tube concentric cylinder and triples tube heat exchanger (TTHX) are different types of heat exchangers used. Better performance is represented by system with longitudinal fins having increasing thermal response of PCM during charging and decrease sub-cooling in discharging. Yosr Allouche et al. [3] (2016) includes the CFD validation for heat transfer in tank with PCM having tubes. The tank with tubes was considered to be horizontal with the datum. The model is validated mathematically for three different rates of flow. The geometry consists of two sides, at one side the charging tubes placed between PCM and on the other side discharging tubes surrounded by PCM. The symmetry exists due to cylindrical shape of tank. At the time of charging heat is transferred from the tubes to PCM and during the time of discharging the heat from PCM transferred to tubes. Low flow rate of three different types were considered to obtain the appreciable variation of temperature at inlet and outlet. A. Allouhi et al. [4] (2018) The paper worked on integrated collector storage solar water heaters (ICSSWH) by optimizing melting and solidification process for the prepared geometry. The study basically includes the improved and efficient design of ICSSWH. External climate data, PCM, mass and flow rate were responsible for overall performance analysis of the system. The analysis includes the study of key operating conditions under real climatic conditions. During the charging and discharging process circulation of PCM takes place due to density difference and dependence on temperature. The variation of PCM phase can be represented by the liquid fraction variation with time. F. Fornarelli et al. [5] (2016) have done analysis of latent heat storage for melting process using CFD modelling defined concentrated solar plants performance. Navier strokes equations were applied for unsteady case to examine the heat transfer process. Boussinesq approximation used to deal with the buoyancy effects during the phase change due to density variation, The whole system was arranged vertically to deal with the natural convection flows. Discretization of system based on three main zones HTF, PIPING, and PCM. The melting process completely dependent on amount of thermal energy absorbed with time. The convective motion also responsible for heat flux rate to PCM due to which time to store heat decreases. Nasiru I Ibrahim et al. [6] (2017) deals with the improvement of heat transfer for TES. Two ways to improve heat transfer optimized geometric configuration and improvement in thermal conductivity. Use of nanoparticles for heat transfer enhances the performance of TES if applied with PCM. As the requirement of thermal energy increases the size of TES also increases. The circular finned and longitudinal finned system were designed and placed between the bottom and top plates of arrangement. Eventually the paper depicts that the heat transfer can be achieved either by increasing area of surface or by increasing the thermal conductivity.

Karunesh kant et al. [7] (2018) includes the heating and cooling of PCM in cyclic manner depending on various variable conditions. To deal with variable conditions on walls of storage sin function was used as variable function under calculation. The melting time of PCM was lesser then the solidification time. The two dimensional analysis can be carried out by using finite element method. The energy can be entered in PCM first as sensible heat then as latent heat at constant melting point. The variable heat flux variation with temperature plots shows isotherms and velocity fields. During melting the velocity component starts increasing then decreases but during solidification the velocity components starts with constant value and then decreases to some value and again becomes constant. A. Sciacovelli et al. [8] (2015) presents the performance maximization of heat storage with efficient fins. The geometry of fins in form of tree was used to enhance the efficiency. The technology involved the numerical methods that study the thermal analysis of system. The radial fins can be arranged with proper gaps to achieve better results. The system was unsteady which needs better design for optimization. For different analysis of orientation of fins, the heat transfer rate was different. The orientation of fins with different angles can be arranged in such a way that the area of fin not affected. N H S Tay et al. [9] (2015) examines the effect of melting of dynamic pattern in PCM tank system. Melting through dynamic way takes less time as compared to natural convection. When the two flow rates become approximately equal then the rate of heat transfer becomes more. A parametric report was directed utilizing the CFD model for dynamic melting. Three unique PCM mass stream rates for both parallel and counter stream were examined. The three mass stream rates of the PCM were set to be half time, equivalent to and multiple times the heat transfer fluid (HTF) mass stream rates. This strategy can't be utilized for examination with dynamic PCM. A parametric report was directed by differing the mass stream rates of the HTF just as PCM. Both parallel and counter streams for PCM and HTF were examined to get optimized results.

W Youssef et al. [10] (2018) investigates the modelling and validation of PCM heat exchanger with CFD as tool. Heat exchanger (HE) having PCM with wired tube design was integrated with indirect solar heat pump system. Some advance technologies were employed to enhance the heat transfer to store heat in large amount. Organic PCM was more stable for experimental purpose. PCM with longitudinal fins balances and installed with graphite. The recreation results were noteworthy to comprehend the working system of PCM and impact of working condition on it. LHTES advances the amount of energy to be stored if applied with PCM as material to store large amount of heat. The spiral design used to enhance the conductivity and rate of heat transfer. Saleh Almsater et al. [11] (2017) conduct the experimental analysis of PCM in vertical triplex heat exchanger (HE) using CFD. To enhance charging/discharging rates in latent thermal energy storage systems (LTESS) tubes with extended surfaces having different orientations used. Geometry was arranged in such a way that PCM is partitioned by fins between the inner and outer portion of heat transfer fluid (HTF). The experiments arranged for both axial and radial directions. Heat rate with time first decreases suddenly and then becomes constant. It is concluded that 86% of energy released by PCM captured by HTF during solidification. The model completely based on natural convection of heat flow techniques. Alberto Pizzolato et al. [12] (2017) worked on the design and optimization of fins from PCM shell and tubes of LTESS. In this the natural convection phenomenon way was considered. Navier strokes equations were applied to solve the transient based situations with porosity change terms included. Due to use of high energy density the size of tank reduces which yields to decrement in cost. LHTES design can be optimized by use of topology concepts. Zhang Jing Zheng et al. [13] (2018) worked on eccentric improvement of flat shell tube latent heat thermal energy storage (LHTES) whose principle depends on the charging/discharging conditions with PCM. The performance of storage with PCM depends on rate of heat transfer. This work relates the heat transfer with the eccentricity between the inner layer and the outer layer of storage. Performance of charging the PCM not always depends on the greater eccentricity, there exist the shortest one on which the performance is better. The melting execution of PCM in flat container was superior then vertical compartment and the solidification execution of PCM in flat container was equivalent to vertical container. The design was completely based on the concept of shifting of center of inner tube with respect to co-centricity of outer tube. For the melting process, there was a minimum value of eccentricity for which melting time was minimum. Karthikeyan Kumarasamy et al. [14] (2017) includes novel CFD based numerical plans for conduction overwhelming embodied PCM phase with temperature hysteresis for hot vitality storage applications. PCM was used in a storage of cylindrical form in the encapsulated manner which increases the thermal response. It was suggested that the CFD based conduction predominant heat source /sink conspire produced for encapsulated phase change material (EPCM) in current investigation ought to be consolidated in to vitality re-enactments. EPCM can be synthesized by applying various known methods so that they can be applied under experiments. EPCM have circular cross section with some depth. For the same EPCM the curves have been compared for solidification and melting time. Basically the variation can be analysed for specific heat EPCM was the most normal approach to coordinate with thermal storage application. This research creates numerical plans to catch the thermal conduct of

EPCM with temperature hysteresis. Monia Chaabane et al. [15] (2014) It includes the study and examination of the performance of integrated collector storage solar water heater (ICSSWH) with PCM. Two numerical methods in three dimensional displaying were created. First depicts a reasonable sensible heat storage unit (SHSU), permitting approval of the numerical model. In light of the great understanding between numerical outcomes and test information from writing and as this kind of sun oriented water radiator introduces the inconvenience of its high night misfortunes, they propose to coordinate a PCM legitimately in the authority and to consider its impact on the ICSSWH heat execution. Based on three residues two PCM layers studied. Sangki Park et al. [16] (2016) includes the examination of movable heat storage system with PCM. For heat storage applications intended to recuperate and reuse squander heat vitality. It was generally beneficial to store heat in a PCM. One dimensional numerical examination and assessment of heat storage framework that utilizes a PCM to store inactive heat. The purpose to design moving heat storage system was to utilize this energy to run vehicles of low energy consumption at night time which were solar operated. There was a big difference between theoretical water energy storage and PCM energy storage. Use of extended surfaces incorporated to increase the heat flux rate. Some sensors and controllers used to examine the proper working under charging/discharging condition of PCM. The experimental model was less efficient than numerical model this was due to ignorance of some practical conditions. Soheila Riahi et al. [17] (2017) examined reverse heat transfer effect on solidification/melting process in LHTES with periodic impact. A numerical report has been directed on shell and tube TES where by the channel heat exchanger (HE) liquid bearing was occasionally turned around amid charging and discharging. Results for the charging forms demonstration on a higher heat exchange territory that creates amid beginning periods and intensification of rectangular convection after soften division. In contrast with the fixed stream condition random stream inversion for the release cases results in an expended heat exchange region for a more drawn out of time. Results shows that as much higher the surface area during heat transfer for particular time period higher the heat flux during charging and discharging. Mikail Temirel et al. [18] (2017) worked on the cooling of PCM with convective heat transfer. The PCM investigated under two conditions one simple and other with nanoparticles. The results on convective cooling compared with the conduction heat transfer condition. The time of discharging was found to be lesser continually due to PCM resistance during convective cooling. Under the natural convection the flow also hindered due to variable density of PCM with temperature change. Most of the experimental studies conducted by taking constant temperature condition but actual results can be found by taking temperature dependent properties. For cooling through convection the heat flux rate for external surface may be higher than the inner surface. The PCM was placed inside the water with adapter, data logger and air speed controller. Thermocouple was used to measure the temperature range. The thermal conductivity increased or exchanged by addition of nanoparticles due to asymmetric and dendritic solidification the time reduces to some extent under natural convective condition.

Changhong Wang et al. [19] (2016) includes the uplift of heat transfer of PCM composite materials. One of the reason to use PCM was its light weight and large specific surface area. For experimental basis to construct a thermo physical properties platform was compulsory. The boundary condition was the most important part of handling with analysis of TES. The heat flow irregularly decreases up to some extent and then increases to basic level. Study can be organized at different points in PCM vertically downward to each other. The boundary conditions were set to be adiabatic at the top and bottom parts. The efficiency of PCM can be studied under with and without existing foam of copper. After the analysis of PCM and PCM with foam the graphs were compared to see their effect on charging conditions. The graph of phase transition for pure paraffin moves lower than the PCM graph for temperature changes. G Diarce et al. [20] (2014) examines the ventilated façade model with PCM using CFD. The PCM applied on outer face only. The study includes the comparison of numerical model with experimental validation. For the analysis the PCM experienced in solid form. The convection and radiation considered under turbulent flow conditions. The vitality proficiency of a ventilated façade was improved when a low temperature diffusivity external layer was utilized. A ventilated dynamic façade containing PCM developed to perform experiments. The reproduced outcomes were thought about with the test results acquired in equivalent period as the model facade. The flow can be studied at various velocities to see the variation of convection heat transfer. PCM, inner side and external layer temperature can be monitored. Transient simulation compared with real experimental data and eventually the simulate time was reduced. The final model was suitable for optimized façade under turbulent condition. N H S Tay et al. [21] (2012) Includes the practical analysis of TES with PCM for tubes. Four tubes placed between the arrangements of cylindrical shape. The examination was to optimize the charging/ discharging time of TES. In the whole process the effectiveness has also balanced with in the reasonable limits. It has been counted that PCM storage had denser energy capacity than sensible heat storage (SHS). The arrangement in cylinder consists of four types of vertical tubes with resistance temperature detectors (RTDs) and thermocouples connections. The experiments at room temperature arranged for freezing process. To analyse the whole process points at three different locations were considered. On which the variation with experimental values compared. It has been found the experimental data matches with the CFD model output data. N H S Tay et al. [22] (2013) examine the difference of using finned with pinned TES with PCM. Three different types of models studied using CFD simulations. All the three models were different in geometric configuration but the concept behind was same. Through these analysis of model, got best design possible for experimental work. These experiments were used to prove whether fins better than pins and vice-versa. Low thermal conductivity of PCM affects the heat transfer rate due to use of extended surface the contact area of heat transfer increases due to which heat flux rate increases. The study of three dimensional model performed on one fourth of the model design to get best possible analysis due to symmetry. The effectiveness should be considered as mean value not local value. The conduction analysis was completely based on freezing process in which over conduction takes place. B Zivkovic et al. [23] (2001) includes the analysis of different shaped containers for isothermal process with PCM. Computational model studied with the mathematical model validation in parallel with the experimental data. The analysis basically focused on the modes of heat transfer. Conduction inside PCM with direction vector of moving fronts. Convection outside with airflow and thermal conductivity of material used. The volume decrement of storage was more in air based systems and less in liquid based systems. Most common method applied for calculation were enthalpy based methods. The temperature first increases linearly and then up to certain value of time it remains constant and then increases parabolic manner. The results show that the experimental data verified the computational data. Sohif Mat et al. [24] (2013) includes the analysis of tube with PCM having fins inside and outside for heat transfer enhancements. The methods were divided in to three parts such as Outer faces, inner faces and both faces of tubes. Due to higher value of energy density the TES with PCM came out to be best way to store solar energy. LHTES shows better performance and advantages with high capacity to store energy. The geometry consists of two co-centric cylinders with fins arranged on inner face, inside portion of outer tube and on both sides in the PCM cavity. The computational study performed on half portion of geometry due to design symmetry. Both peaks of temperature during solidifying and melting time were of same nature and following the same pattern. As the number of fins increases the heat transfer surface increases due to which heat flux rate also increases. Weihuan Zhao et al. [25] (2013) studied the PCM in encapsulated form and analyze the heat transfer. TES seems to be the perfect replacement of source of energy during off hours. TES based on the concept of storing heat in the form of latent heat and sensible heat by PCM material. The time of charging, storing heat and discharging depends on thermal conductivity and type of PCM used. Sometimes heat also stored in chemical energy storage governed by chemical reactions. Generally organic PCM have lower value of thermal conductivity and energy density hence researchers moved for inorganic PCM, but they are less stable. The thickness of PCM layer depends on the critical value of cylindrical radius. The time for melting also depends on the size of encapsulated PCM particles. Mohammad Reza Kargar et al. [26] (2018) includes the examination of TES with PCM for direct parabolic trough solar power plants (PTSPS). The cascade arrangement consists of preheater, generator, regulator etc. The numerical model validated with experimental data. Various parameters investigated for best possible design. The irregular pattern of solar energy was the key to unlock the concept of TES with PCM. The performance of energy storage depends on charging/discharging time of PCM and its optimization. The area closest to the flow steam melts first and the melting propagates from the inner to the outer layer. The first analysis includes the different parameters which effects output steam, PCM. Secondly the effect on preheater and super heater analyzed. Results also include the analysis of best possible available energy called as exergy. The graphical trends were approximately close to the experimental data investigated. The basic concept of the paper was to optimize the performance of latent heat thermal energy storage (LHTES) system which was directed by the steam generated from solar energy usage. The performance apart from running conditions also depends on the properties of the material and PCM used. C Y Zhao et al. [27] (2010) worked on the enhancement of thermal energy storage (TES) with phase change material (PCM) having metal foams. Metal foams are added in the PCM to enhance its properties of energy density and thermal conductivity. During the melting phase of PCM the convection increases rate of heat dissipation naturally. The metal foams addition increases the overall heat transfer coefficient. The metal foams with PCM decreases the solidification time by increasing solidification rate. Thermocouples, data logger, voltage transformer, power inputs were attached with the system to control working, store the performance parameters and analyze the various recorded values. The graphical trends of temperature with time for PCM with foam were higher than trends for PCM alone. The motion of melting fronts from liquid phase of PCM to it solid phase was considered to be uniformly dissipated with directional vector. The lesser value of porosity and density of pores can increase the heat transfer coefficient due to which heat flux rate increases. The process of discharging can be investigated under natural cooling conditions.

**2. Computational Domain for Models**

This investigation is conducted for improvement of design of thermal energy storage (TES) with gallium as phase change material (PCM). The shape of the design is in the form of cylindrical cavity. The complete study carried out in 2- dimensional analysis on the software of computational fluid dynamics (CFD) named ANSYS 14.5 version. For the complete analysis to obtain optimized design of TES, this investigation worked on 5 models of different designs. The simulation of all these models under same working conditions analyzed on Ansys fluent software. The conditions provided to the analysis was same to get the difference in results, this clearly shows which design is better according to working conditions and the saving of material during optimization of design also considered. Initially model-1 was designed in the form of cylindrical cavity with four heat source and all the other models compared with this initial design and the difference according to working conditions noticed. The gallium as PCM is used for the analysis of all models. The gallium has the advantage of low melting point with 29.7°C and has high thermal conductivity of 32W/m-K. Generally, PCM have very less thermal conductivity (<0.2W/m-K), compared to that gallium has good range of thermal conductivity. This increases the rate of heat flux in to and out of the PCM body. The time required for melting is less to store same amount of heat as compared to other PCMs. The inner walls of heat source were considered to be made of aluminium. The outer wall of cylindrical cavity is adiabatic in nature. The outer wall assumed at constant atmospheric temperature but zero heat flux was considered. There are total five models designed in 2D which are given below. All the models with fins on heat source has constant surface area of each heat sources of that model. The total surface area of all heat sources with fins is constant = 507.32 mm².

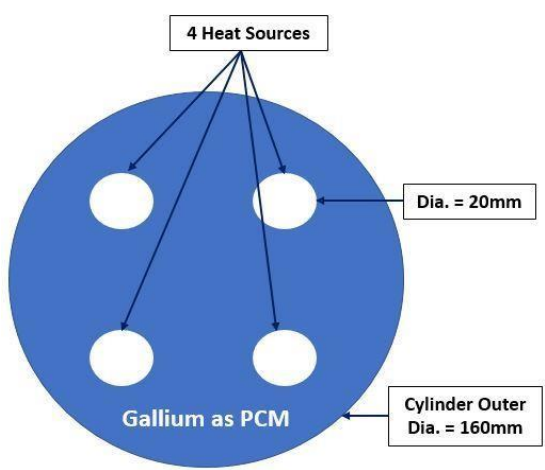 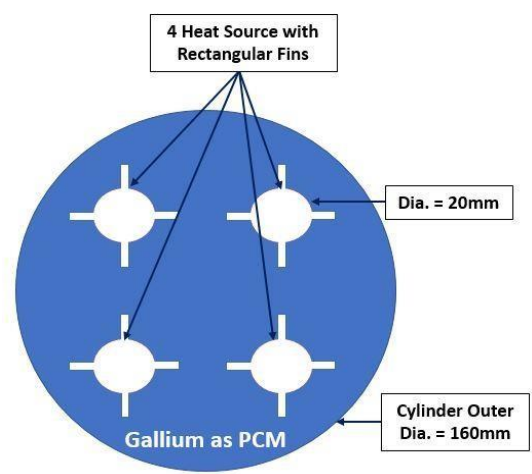

**Fig. 1.** Computational Domain of Model 1     **Fig. 2.** Computational Domain of Model 2

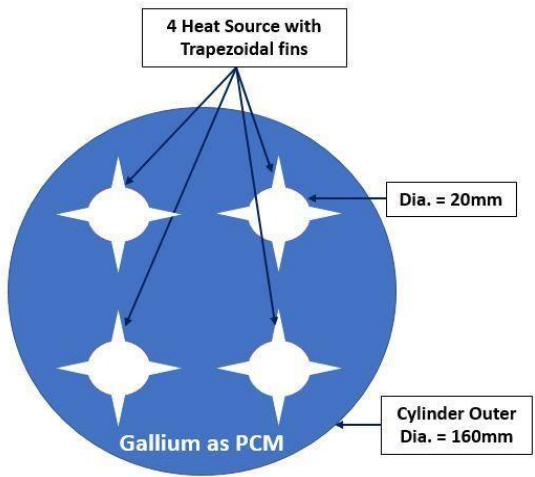

**Fig. 3.** Computational Domain of Model 3

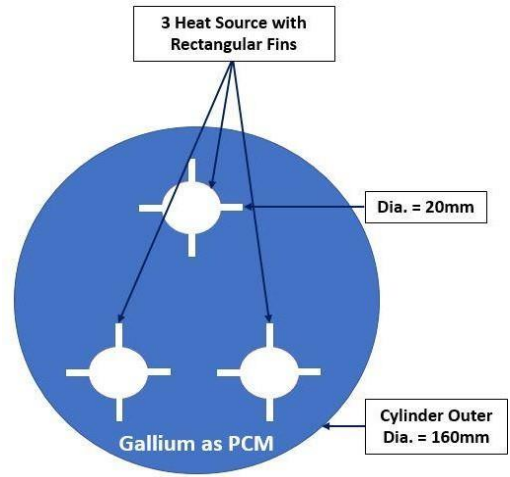

**Fig. 4.** Computational Domain of Model 4

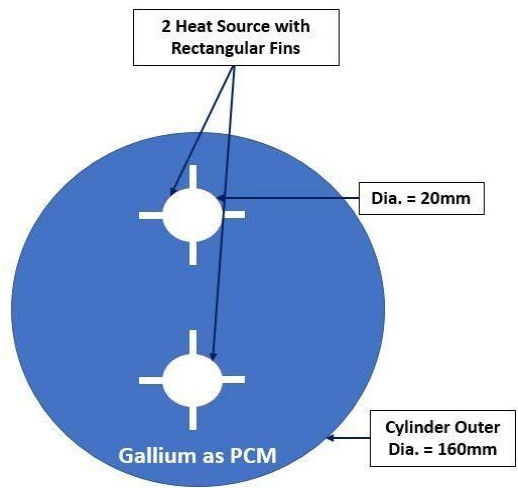

**Fig. 5.** Computational Domain of Model 5

**Table 1. Geometrical parameters of all models**

| Description | Model 1 | Model 2 | Model 3 | Model 4 | Model 5 |
| --- | --- | --- | --- | --- | --- |
| Outer diameter of cylinder, (mm) | 160 | 160 | 160 | 160 | 160 |
| Diameter of heat source, (mm) | 20 | 20 | 20 | 20 | 20 |
| Number of heat source | 04 | 04 | 04 | 03 | 02 |
| Length of each fin, (mm) | 7 | 7 | 7 | 12.3 | 21.3 |
| Breadth of each fin, (mm) | 2 | 2 | Base dimension: 3 Upper dimension: 1 | 2 | 5 |
| Surface area of each heat source, (mm²) | 62.832 | 126.83 | 126.83 | 169.106 | 126.83 |
| Perimeter of fin, (mm) | 16 | 16 | | 26.6 | 47.7 |
| Total no. of fins | 0 | 16 | 16 | 12 | 08 |
| Angular Orientation of heat source | 90° | 90° | 90° | 120° | 180° |

## 3. Boundary Conditions

This study includes the design of five different models. All models have different geometries. The 'Model 1' don't have any fins in its geometry and all the rest four models have fins of different shape and size. The designs were compared with the help of CFD simulations on ANSYS 14.5 version in FLUENT module. The operating and boundary conditions for all models were exactly similar. The temperature of heat source and fins assumed to be constant 343K for complete analysis. The outer walls of cylindrical TES assumed to be adiabatic and at atmospheric temperature of 298K.

**Table 2**. Applicable Boundary conditions for all models.

| Parameter | Values |
| --- | --- |
| Heat Source Temperature | 343K |
| Outer walls of cylindrical TES | Adiabatic & Maintain at 298K |
| Phase Change Material | Gallium |

### 3.1. Grid Generation

Improvement of results can be determined by the use of smaller cell sizes for calculation, determined by grid sensitivity. It reflects the criterion of becoming finer mesh to generate accurate results. Initially the CFD meshing starts with coarse size mesh then moved towards refined mesh with time. As much the mesh is finer the results of post processing are more accurate. But there is a limit to refinement also otherwise the back flow of fluid occurs. The conditions for steady state criterion are-

- Residual values range ($10^{-4}$ or $10^{-6}$).
- Monitor points reaches to stability.
- Imbalances $< 1\%$.

The results obtained from simulation depends on the number of iterations run under the analysis. The number of iterations completely depends on-

- Size of time step.
- Iterations per time step.
- Number of steps.

When the values start repeating or reaches to a steady value and the domain overall imbalance becomes $< 1\%$, Then we monitor these values of simulations and obtained results from there.

### 3.2. Grid independence test

Meshing in ANSYS is a high performance product, the foundation of best simulation. Grids generated during meshing, in forms of cells/zones should be accurate. The accuracy can be checked for each design by performing grid independency test. The best mesh can be obtained by checking the value of same parameter at a point in the design for different number of elements and nodes. In this investigation, all the models check for temperature as the parameter. The number of elements changes with the change in mesh. The value of mesh from where the checking parameter becomes almost steady, gives approximately the same value that mesh can be considered as the best mesh for design. The meshing of all the five models are presented below in Fig. 6. All the five models in this paper checked by performing the grid independency test and the details are given below in Table 3: -

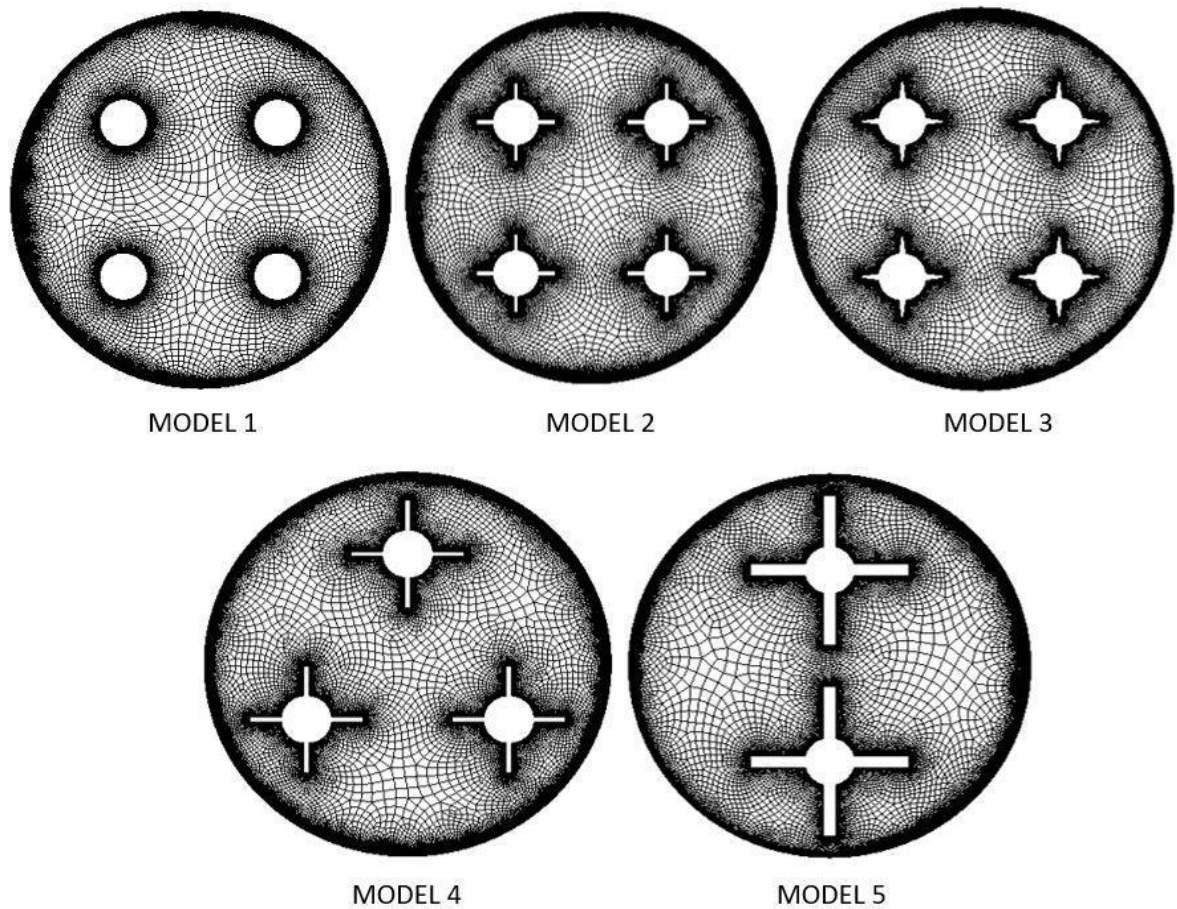

**Fig. 6.** Meshes of All Models

**Table 3.** Details of grid independence test

| Details | Model 1 | Model 2 | Model 3 | Model 4 | Model 5 |
| --- | --- | --- | --- | --- | --- |
| Number of nodes. | 69244 | 51139 | 57786 | 61224 | 43868 |
| Number of elements. | 64748 | 48692 | 55140 | 58631 | 41812 |
| Skewness (<0.95) | 0.21781 | 0.19717 | 0.17928 | 0.16459 | 0.17155 |
| Aspect ratio Avg. value | 1.5338 | 1.56767 | 1.49967 | 1.57035 | 1.56077 |
| Orthogonal quality (>0.01) Avg. value | 0.945269 | 0.94724 | 0.95569 | 0.95775 | 0.95507 |

## 4. General equation:

The generalized equation that covers the conservation of mass, momentum, energy, species. This solver equation depends on finite volume method.

$$\frac{d}{dt}\int_v \rho\phi \partial v + \oint_A \rho\phi v \cdot \partial A = \oint_A \Gamma_\phi \nabla\phi \cdot \partial A + \int_v S_\phi \cdot \partial V \qquad (i)$$

### 4.1. Continuity equation:

$$\frac{d\rho}{dt} + \nabla \cdot \rho\bar{u} = 0 \qquad (ii)$$

$\bar{u}$ – velocity vector, $\rho$ – density.

### 4.2. Momentum equation:

$$\frac{d\rho\bar{u}}{dt} + \nabla \cdot (\rho\bar{u}\bar{u}) = -\nabla p + \nabla r + \rho\beta(T - T_m)g \qquad (iii)$$

Where p- static pressure, T- temperature, $T_m$- mean temperature, $\beta$- coefficient of volumetric thermal expansion, g- acceleration due to gravity vector, $r$- viscous stress tensor for Newtonian fluid.

$$\tau = \mu\,(\nabla u + (\nabla u)^T) \qquad (iv)$$

$\mu$ – dynamic viscosity.

### 4.3. Energy equation:

Enthalpy of material:

$$H = h + \Delta H \qquad (v)$$

Where, h- sensible enthalpy, $\Delta H$- latent heat.

$$h = h_{ref} + \int_{T_{ref}}^{T} C_p \partial T \qquad (vi)$$

$h_{ref}$ – Reference enthalpy, $T_{ref}$ – reference temperature, $C_p$ – specific heat at constant pressure. The above described mathematical equations are the governing equations of the CFD. The equations involved the balancing of mass, momentum and energy of the complete system under consideration. The value of liquid fraction 'α' varies with the temperature, as the temperature increases melting of material increases and the value of liquid fraction also increases with time.

- α = 0, T<$T_s$
- α=1, T>$T_l$
- at particular temperature 'T', $\alpha = \frac{T-T_s}{T_l-T_s}$   where $T_s$ < T < $T_l$

Latent heat content in terms of latent heat of material ($L_f$) can be described as:

$$\Delta H = \alpha\, L_f \qquad (vii)$$

The latent heat content can be ranged from '0' for solid to '$L_f$' for liquid.

## 5. Results and Discussion

### 5.1. For Model 1

The geometry of model 1 includes the heat source without fins. The dimensions of heat source pipes and the outer cylindrical shape remains same.

#### 5.1.1. Liquid Fraction Contours

The contours of liquid fraction variation with time for Model 1 as shown in Fig. 7 below-

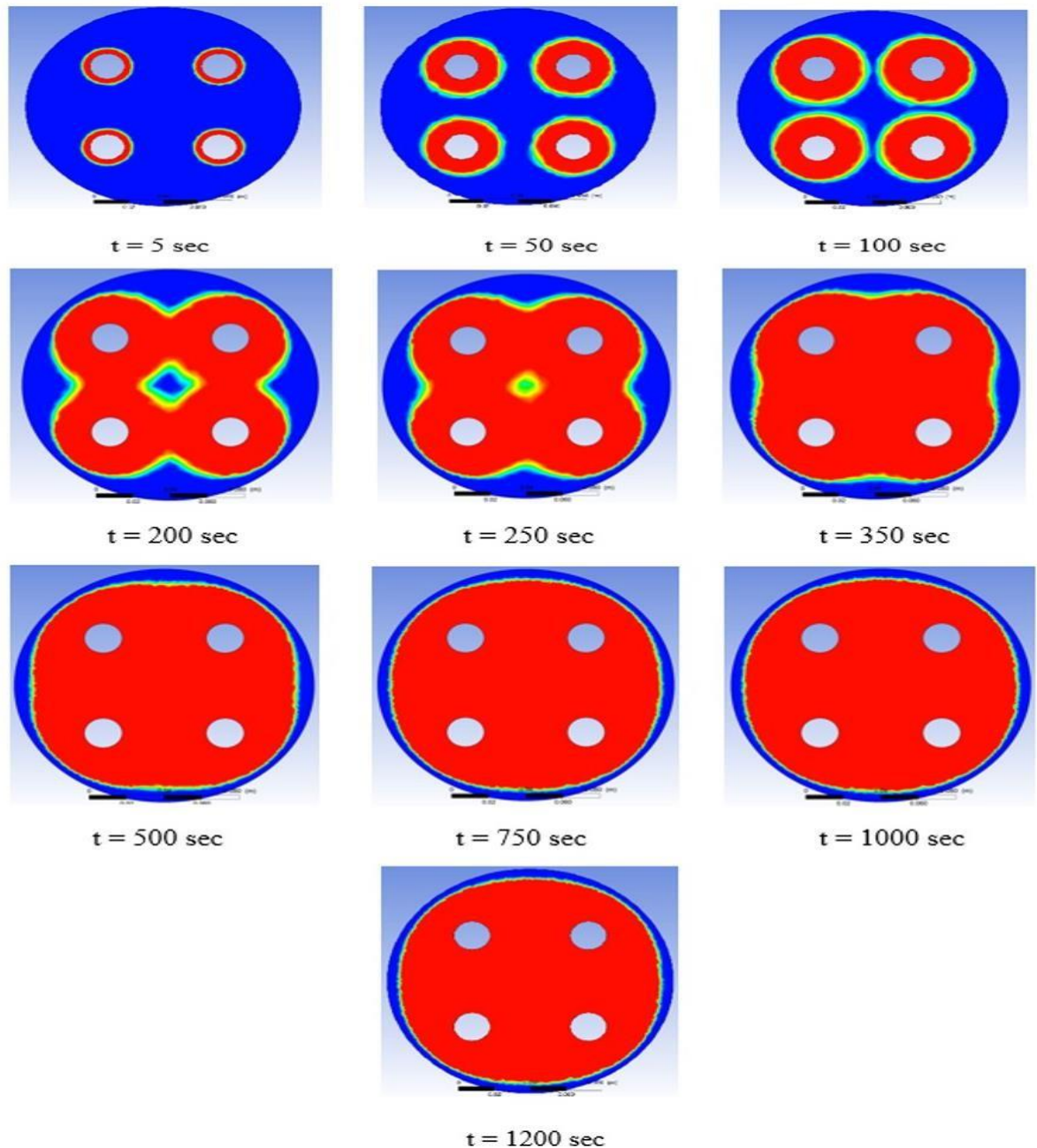

**Fig. 7.** Liquid Fraction Contours variation with time for Model 1

✓ Variation of liquid fraction can be represented on graph as given blow:

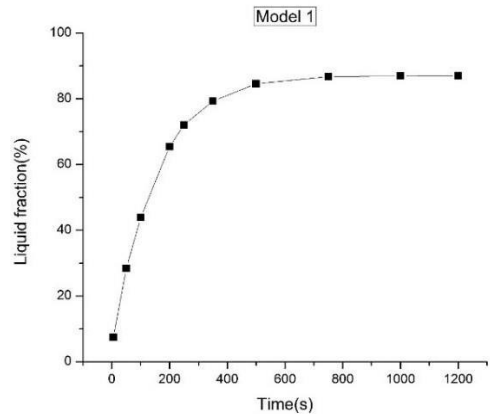

**Fig. 8.** Plot of Liquid fraction variation with time for Model 1

- The contours of liquid fraction for the geometry without fins clearly shows the variation of melting as the time increases.
- The time variation starts from 5 seconds to 1200 seconds.
- At the end the liquid fraction variation becomes almost constant at about 90%.
- Further increase in temperature with time does not affect much on liquid fraction of PCM.
- This clearly shows the variation of liquid fraction with time is non-linear in nature.
- Due to the adiabatic nature of outer walls of cylindrical TES which is maintained at constant atmospheric temperature, the liquid fraction becomes almost stationary at 90%.

**5.1.2. Temperature**

The plot & contours of temperature variation with time presented below in Fig 9 & 10:

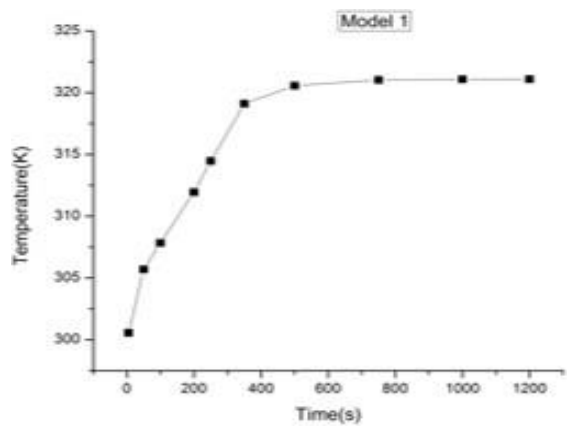

**Fig. 9.** Plot of Variation of Mean Temperature with Time for Model 1

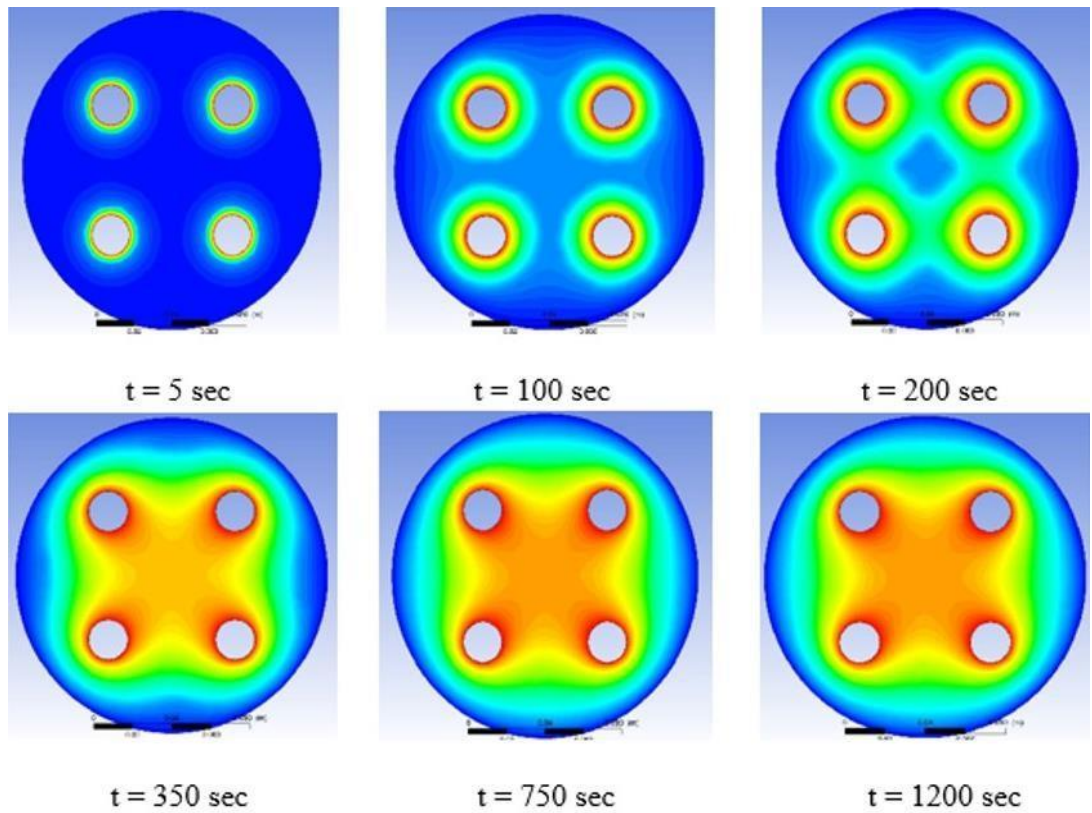

**Fig. 10.** Contours of Mean Temperature with Time for Model 1

### 5.1.3. Velocity (streamlines)

The contours of velocity variation with time for Model 1 presented below in Fig. 11.

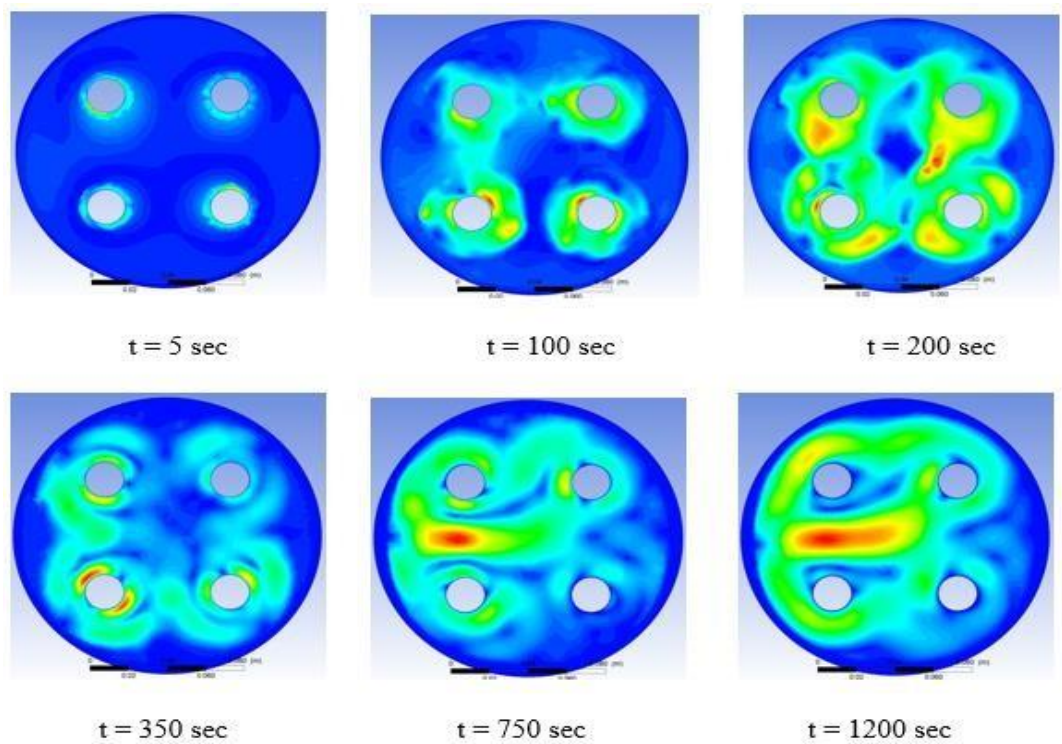

**Fig. 11.** Velocity (streamlines) contours for Model 1

- These contours show the variation of streamlines, velocity vector or velocity of flow of gallium in the body during melting.
- The circulation occurs due to change in density as the temperature increases with time.
- The contours varying from 5 seconds to 1200 seconds clearly shows the variation of streamlines with circulation in all the directions.

### 5.2. For Model 2

The model 2 includes the fins on the surface of heat source. The fins are rectangular in shape. The variation of liquid fraction, temperature and velocity inform of contours shown below:

#### 5.2.1. Liquid Fraction

The contours & plot of liquid fraction variation with time for Model 2 as shown below in Fig. 12 & 13:

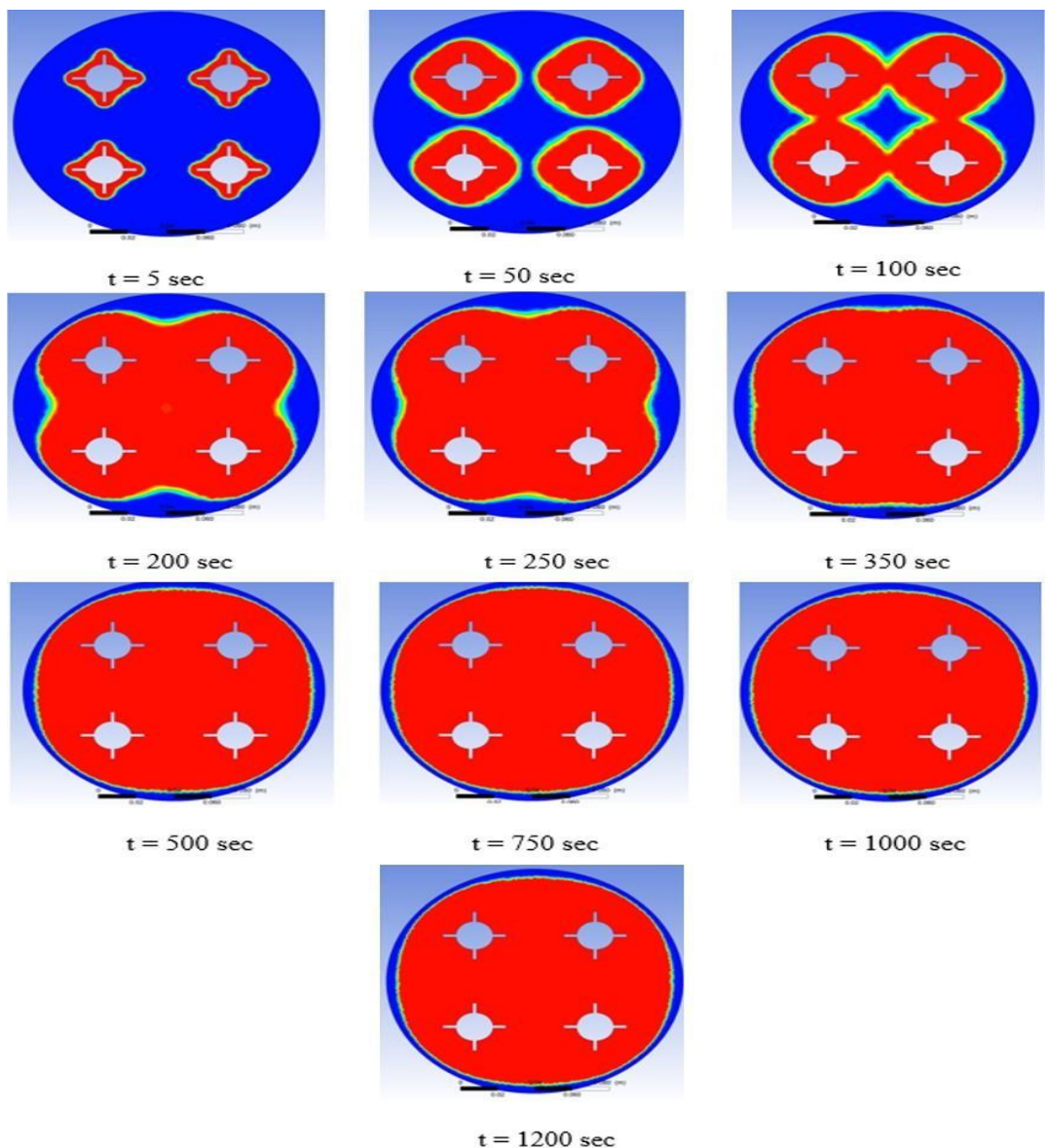

**Fig. 12.** Liquid Fraction Contours variation with time for Model 2

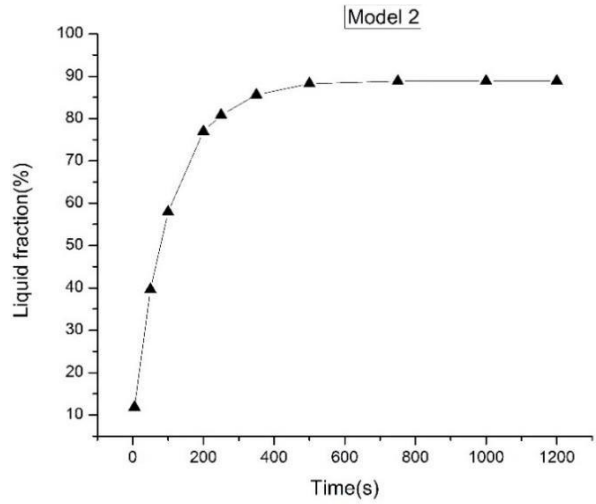

**Fig. 13.** Plot of Liquid fraction variation with time for Model 2

### 5.2.2. Temperature

The contours & plot of mean temperature variation with time presented below in Fig 14 & 15:

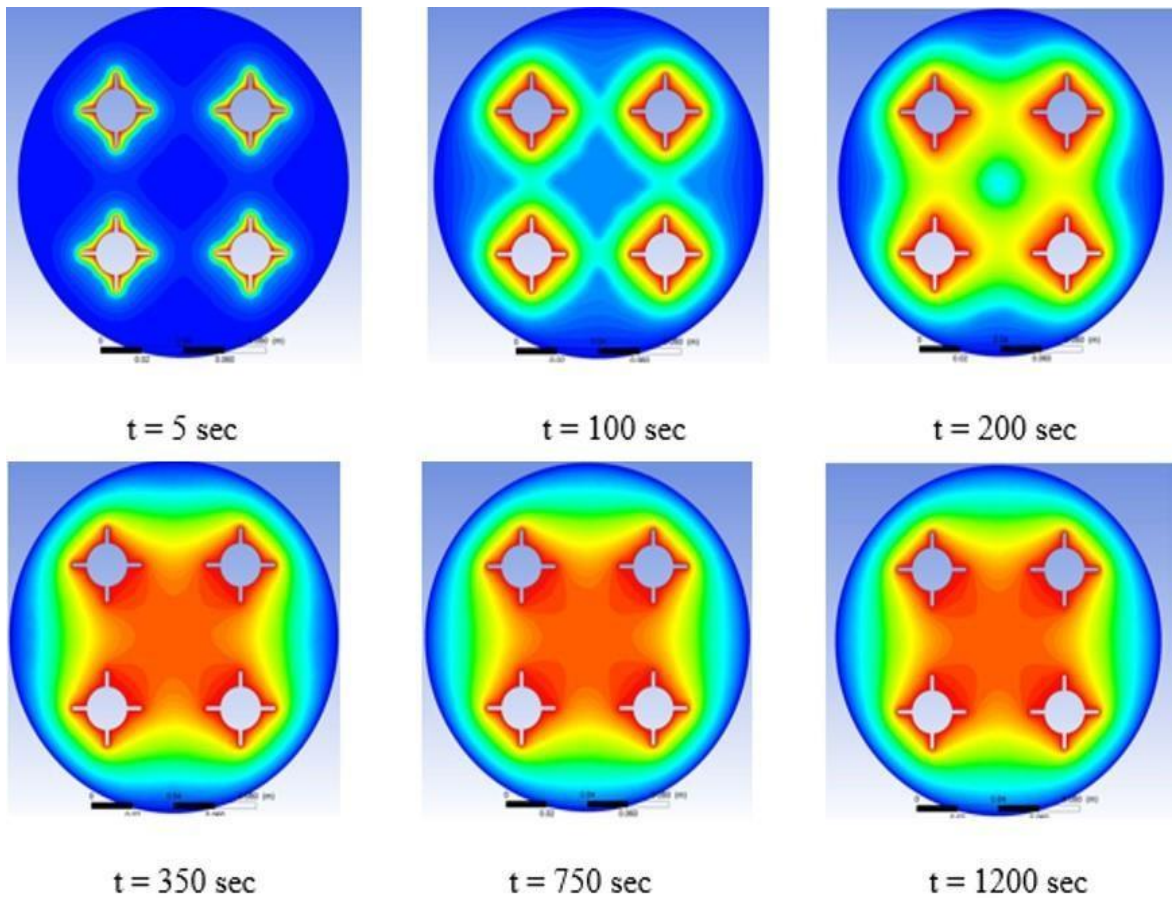

**Fig. 14.** Contours of Mean Temperature with Time for Model 2

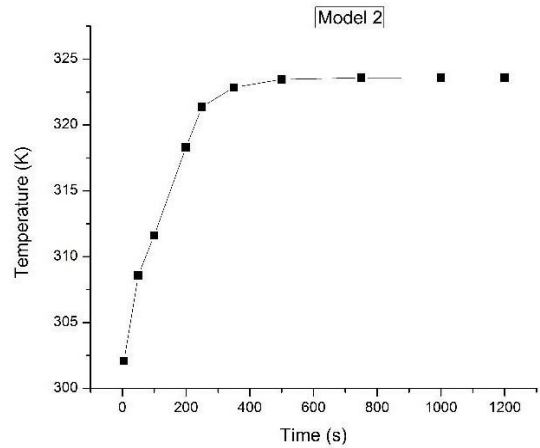

**Fig. 15.** Variation of Mean Temperature with Time for Model 2

### 5.2.3. Velocity (Streamlines)

The contours of velocity variation with time for Model 2 presented below in Fig. 16.

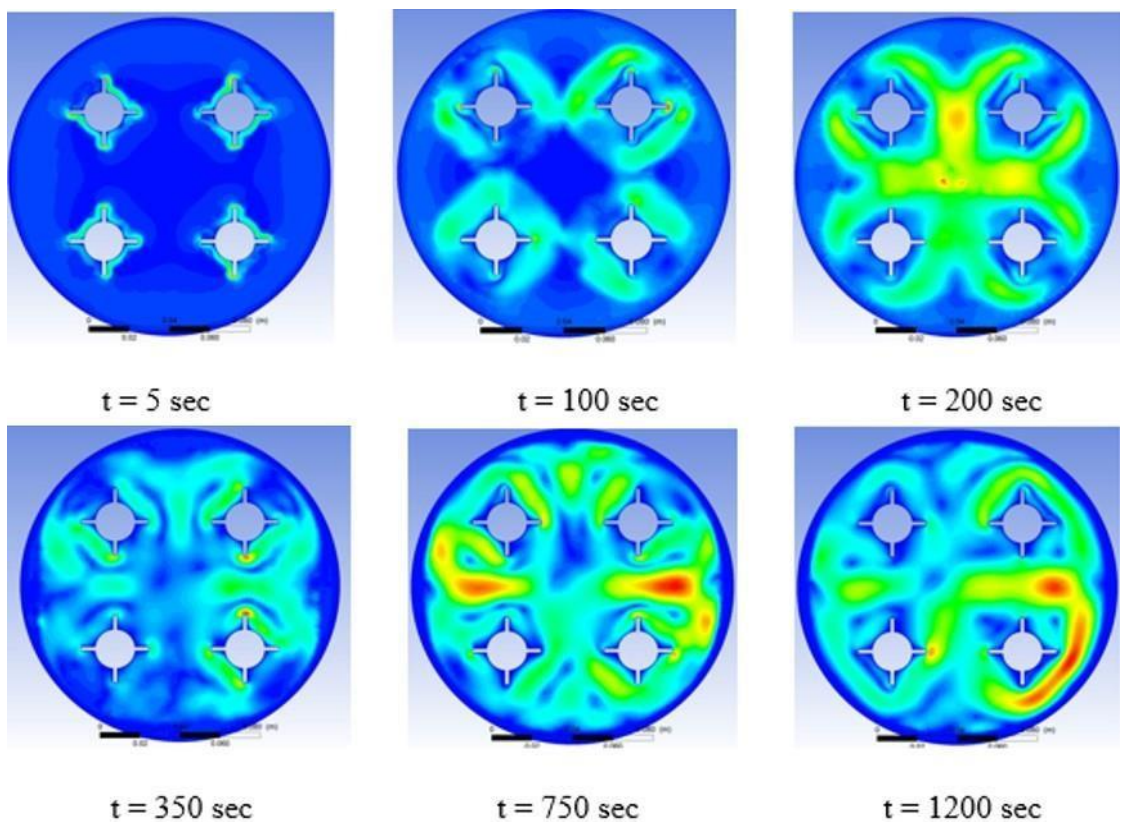

**Fig. 16.** Velocity (streamlines) contours for Model 2

- The contours of liquid fraction clearly show the melting of PCM.
- The melting region moves forward from inner heat source to outer boundary with faster rate as compared to 'Model 1'.

- The streamlines start from inner heat source and moves in outward direction with velocity vectors covering all the directions as shown in contours.
- Due to increase of surface area to transfer heat, the directional heat flux rate also increases.
- The value of mean temperature for this model is maximum at 1200 seconds.

### 5.3. For Model 3

The model 3 consists of fins on the surface of heat source. The fins are of trapezoidal shape. The variation of change in shape can be examined.

### 5.3.1. Liquid Fraction

The contours & plot of liquid fraction variation with time for Model 3 as shown below in Fig. 17 & 18:

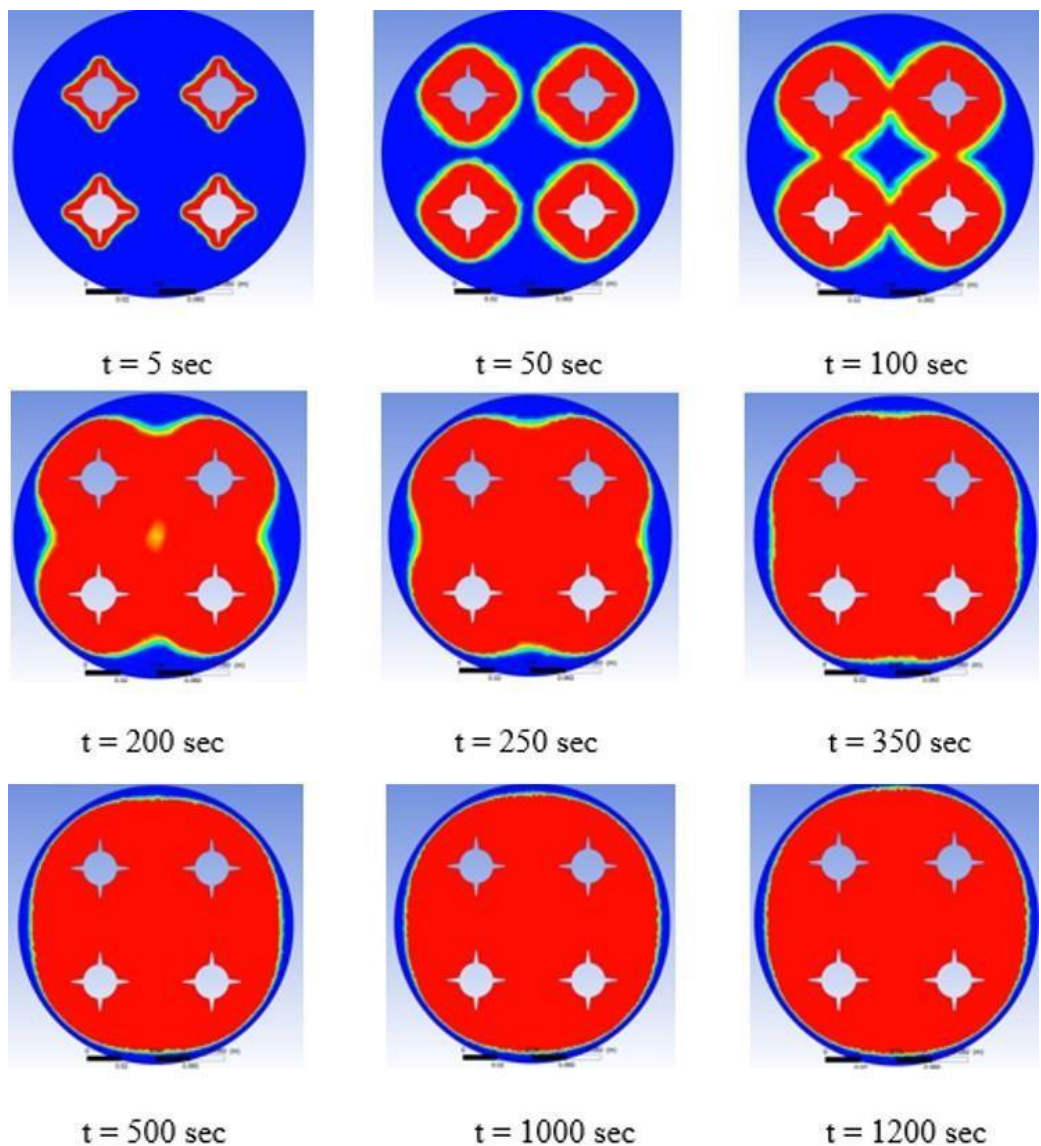

**Fig. 17.** Liquid Fraction Contours variation with time for Model 3

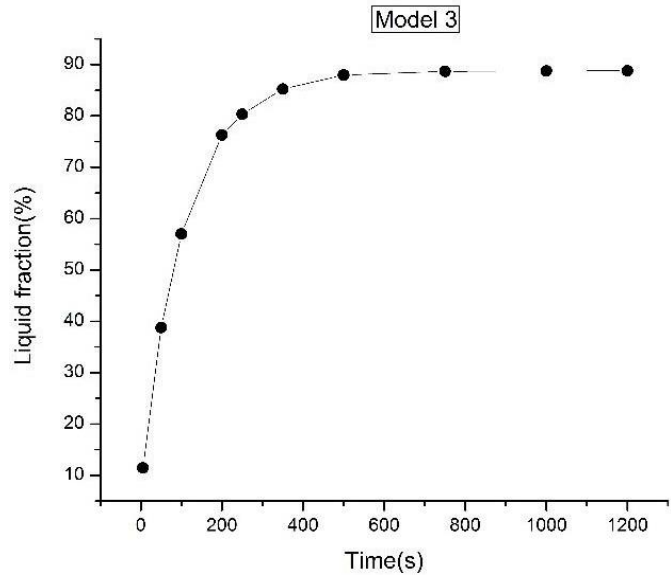

**Fig. 18.** Variation of liquid fraction with time for Model 3

### 5.3.2. Temperature

The contours & plot of mean temperature variation with time presented below in Fig 19 & 20:

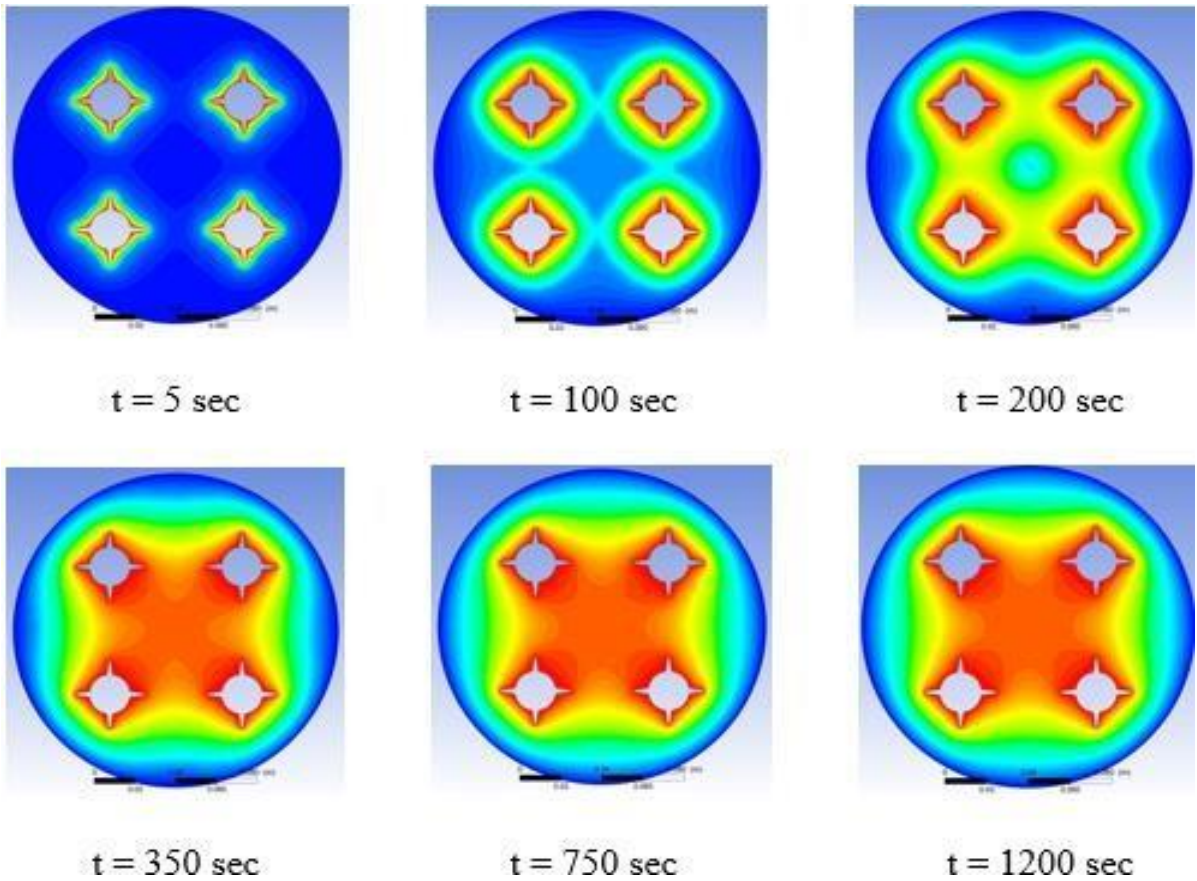

**Fig. 19.** Contours of Mean Temperature with Time for Model 3

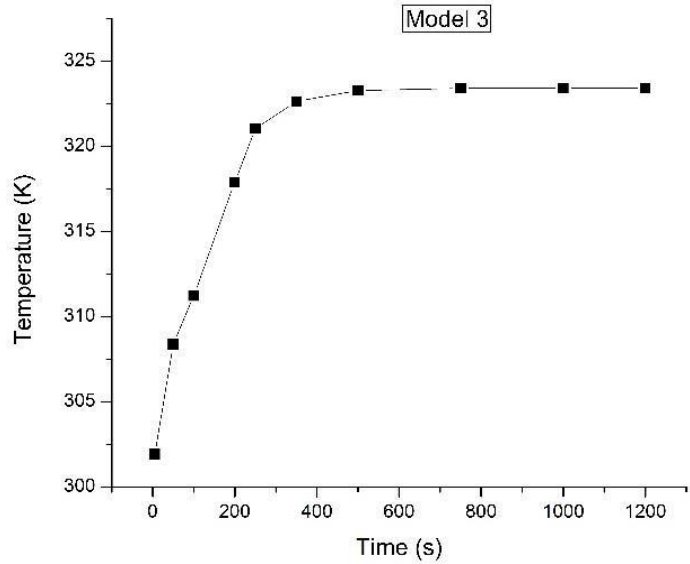

**Fig. 20.** Variation of Mean Temperature with Time for Model 3

### 5.3.3. Velocity (Streamlines)

The contours of velocity variation with time for Model 3 presented below in Fig. 21

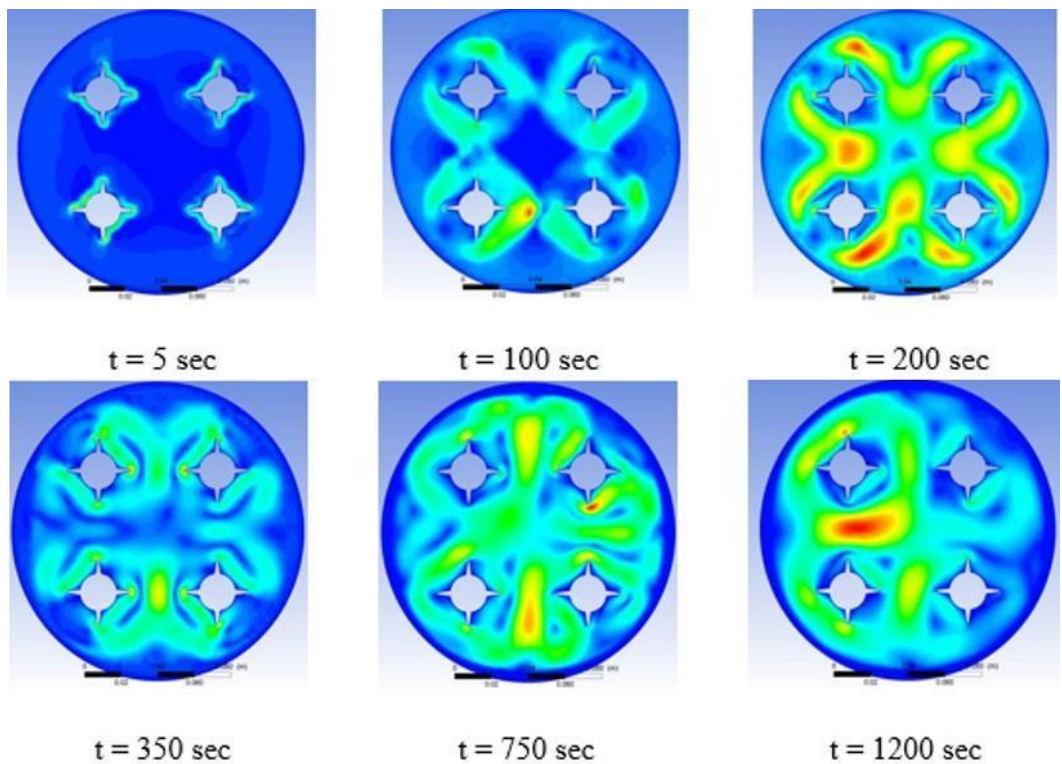

**Fig. 21.** Velocity (streamlines) contours for Model 3

### 5.4. For Model 4

The model 4 consists of three heat sources with rectangular fins. The size of fins selected in such a way that the total surface area of all three heat sources is equal to other models.

## 5.4.1. Liquid Fraction

The contours & plot of liquid fraction variation with time for Model 4 as shown below in Fig. 22 & 23:

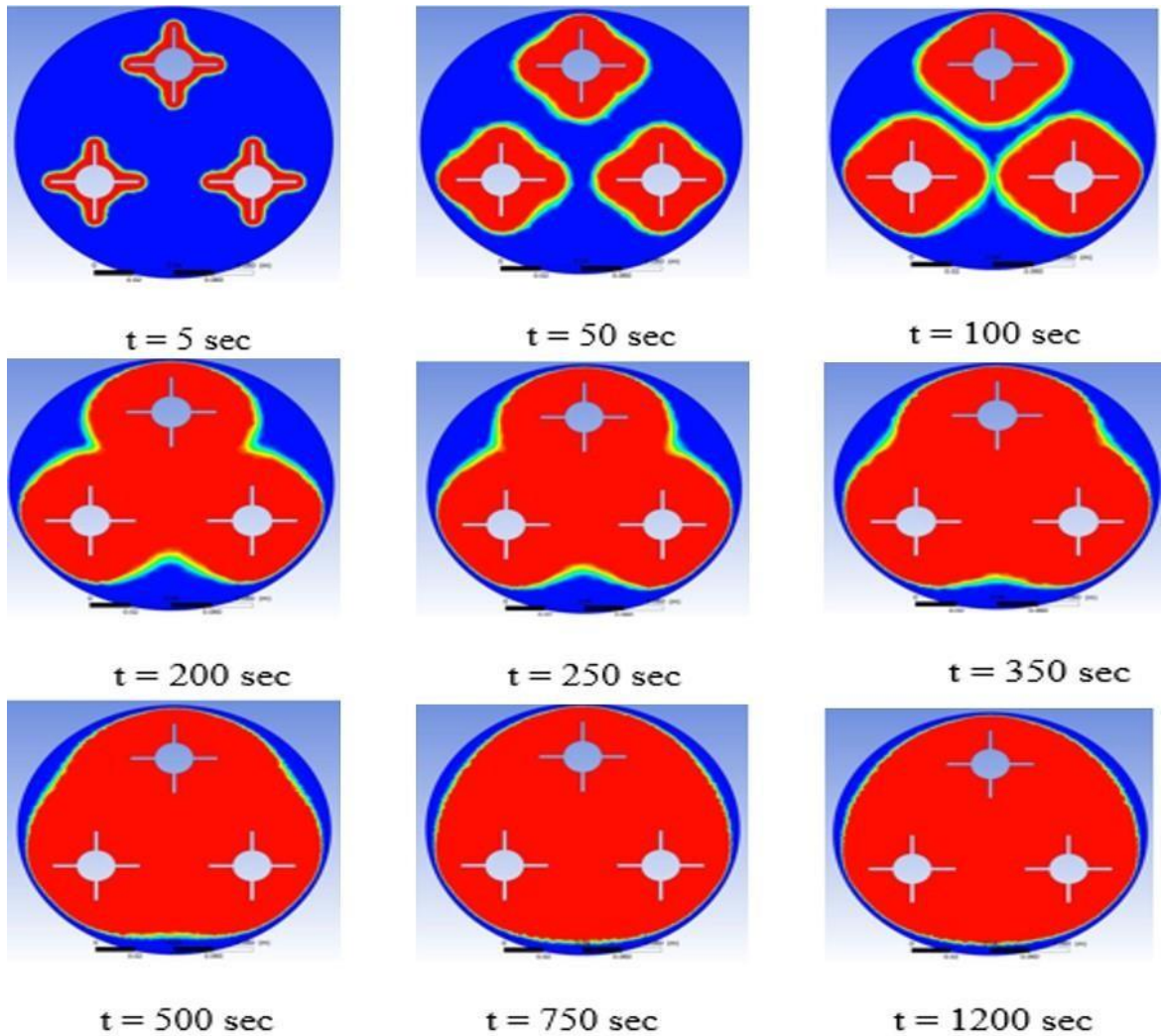

**Fig. 22.** Liquid Fraction contours variation with time for Model 4

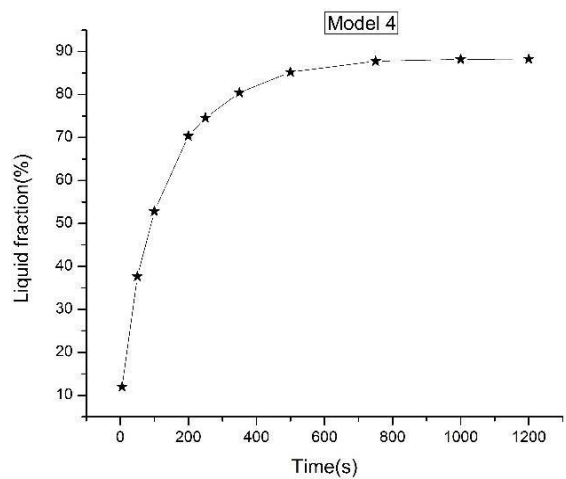

**Fig. 23.** Variation of Liquid Fraction with time for Model 4

## 5.4.2. Temperature

The contours & plot of mean temperature variation with time presented below in Fig 24 & 25:

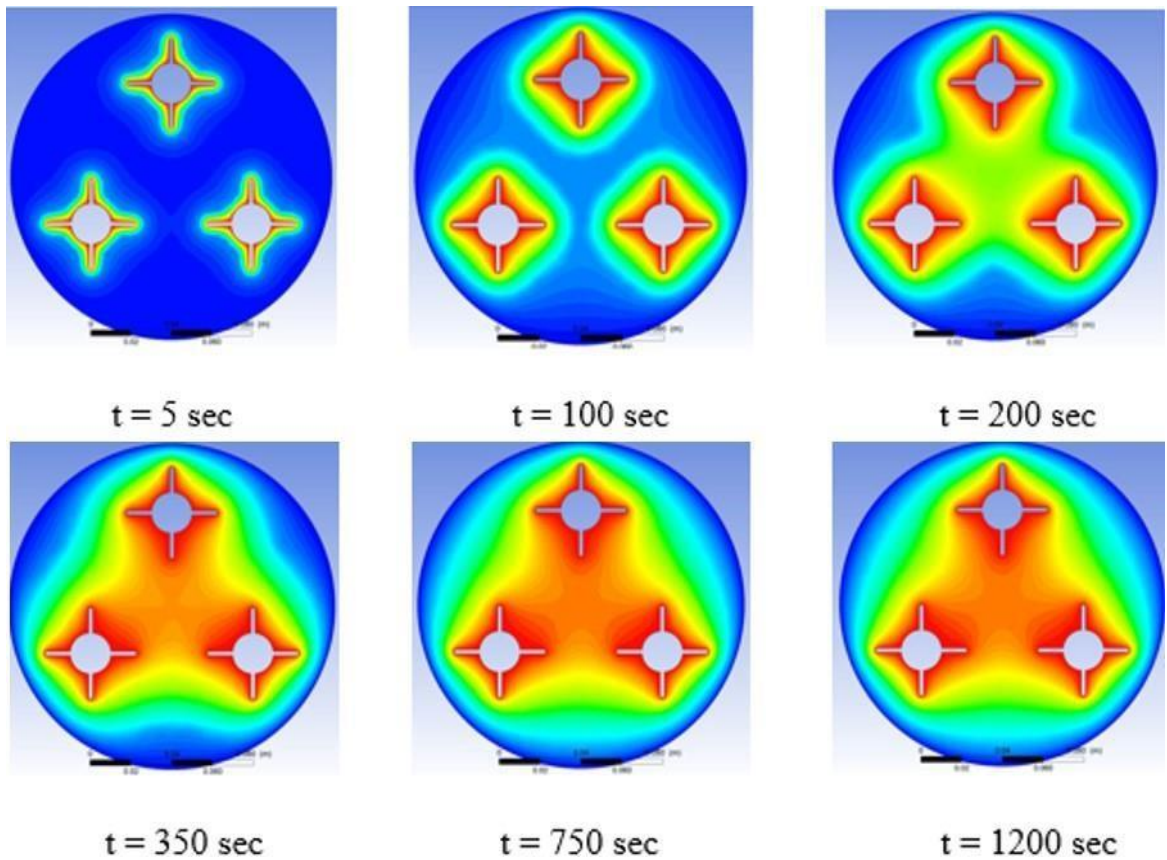

**Fig. 24.** Contours of Mean Temperature with Time for Model 4

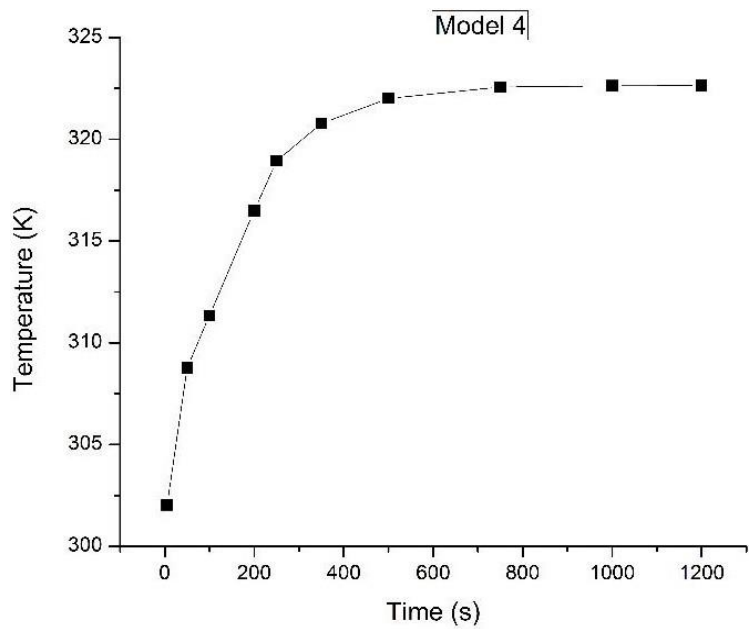

**Fig. 25.** Variation of mean temperature with time for Model 4

### 5.4.3. Velocity (streamlines)

The contours of velocity variation with time for Model 4 presented below in Fig. 26.

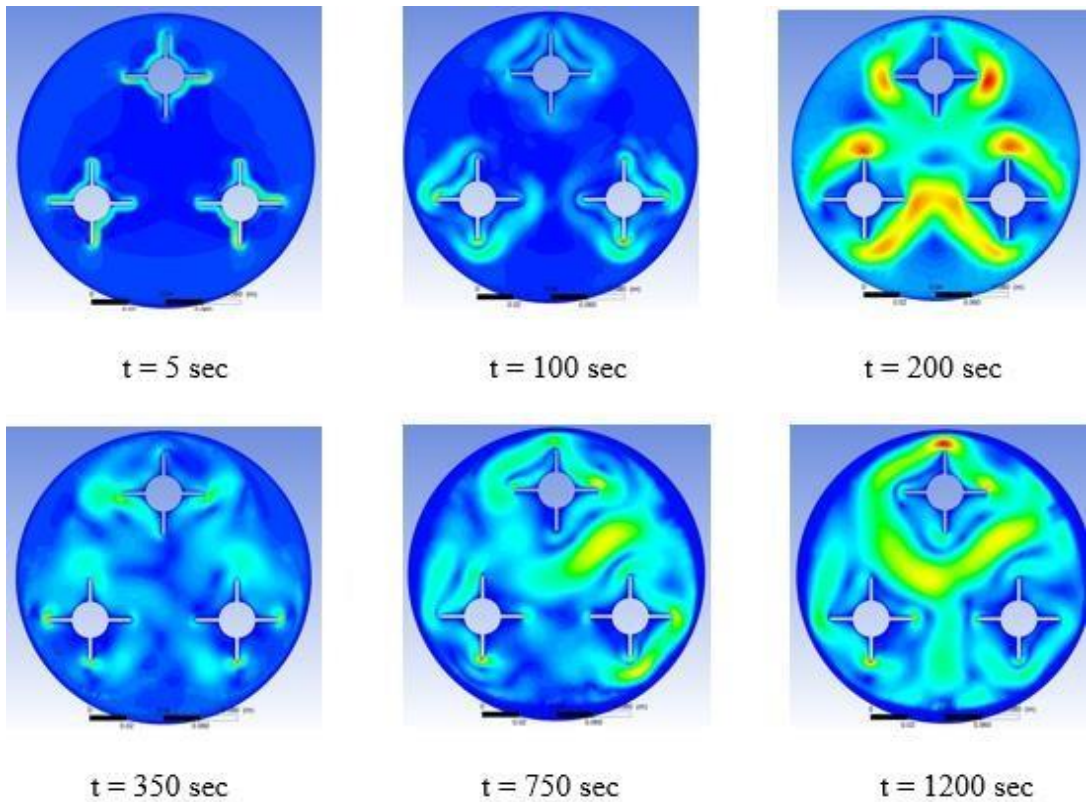

**Fig. 26.** Velocity (streamlines) contours for Model 4

## 5.5. For Model 5

The design of model 5 consists of two heat sources with rectangular fins. The dimensions of fins selected in such a way that the total surface area of heat sources remains constant.

### 5.5.1. Liquid Fraction

The plot & contours of liquid fraction variation with time for Model 5 as shown below in Fig. 27 & 28:

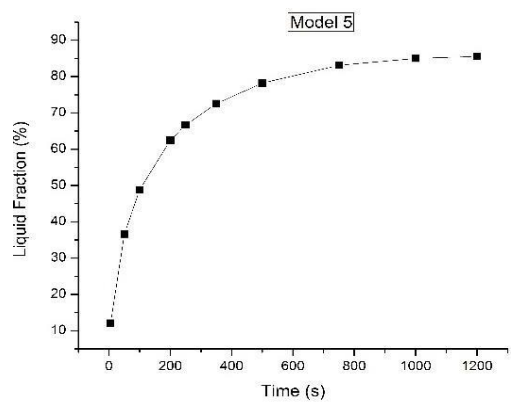

**Fig. 27.** Variation of liquid fraction with time for Model 5

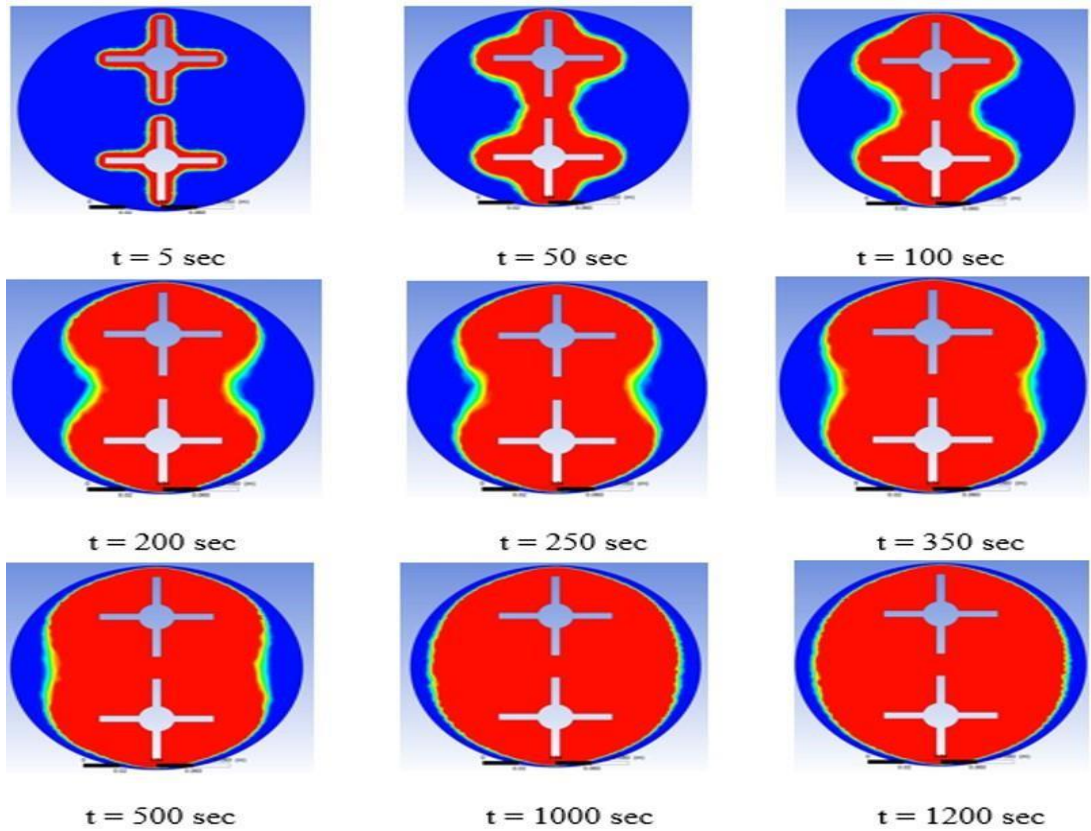

**Fig. 28.** Liquid Fraction contours variation with time for Model 5

### 5.5.2. Temperature

The contours & plot of mean temperature variation with time presented below in Fig 29 & 30:

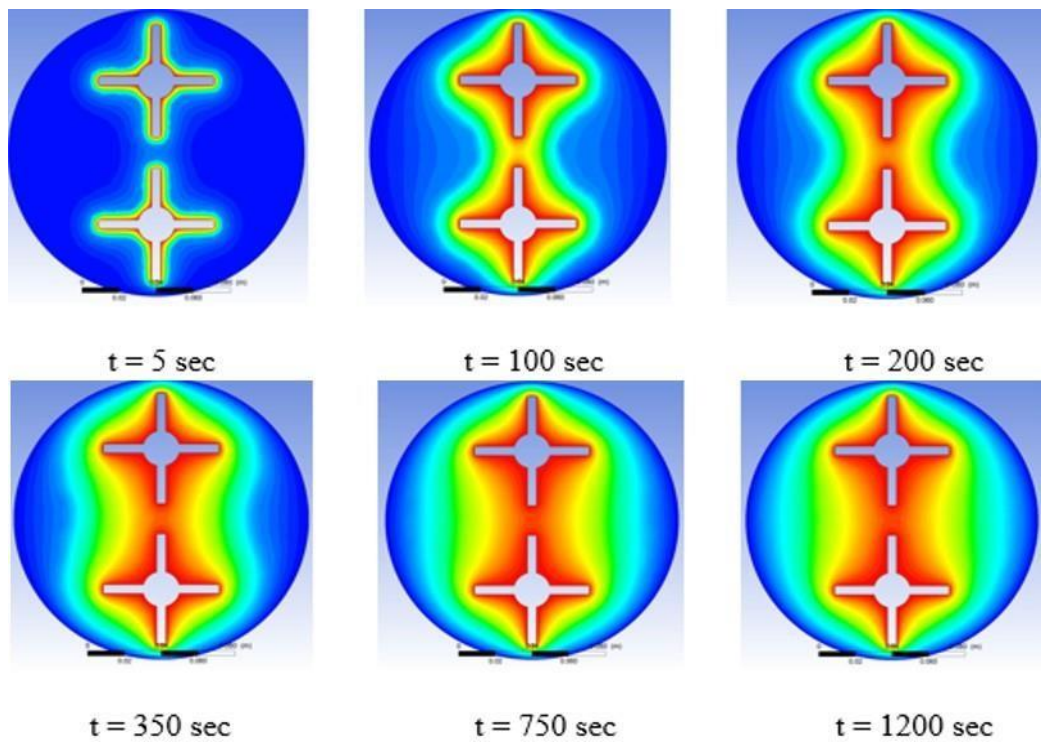

**Fig. 29.** Contours of Mean Temperature with Time for for Model 5

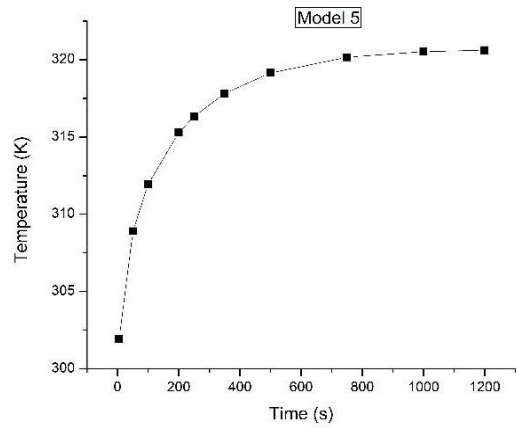

**Fig. 30.** Variation of Mean Temperature with Time for Model 5

### 5.5.3. Velocity (streamlines)

The contours of velocity variation with time for Model 4 presented below in Fig. 31.

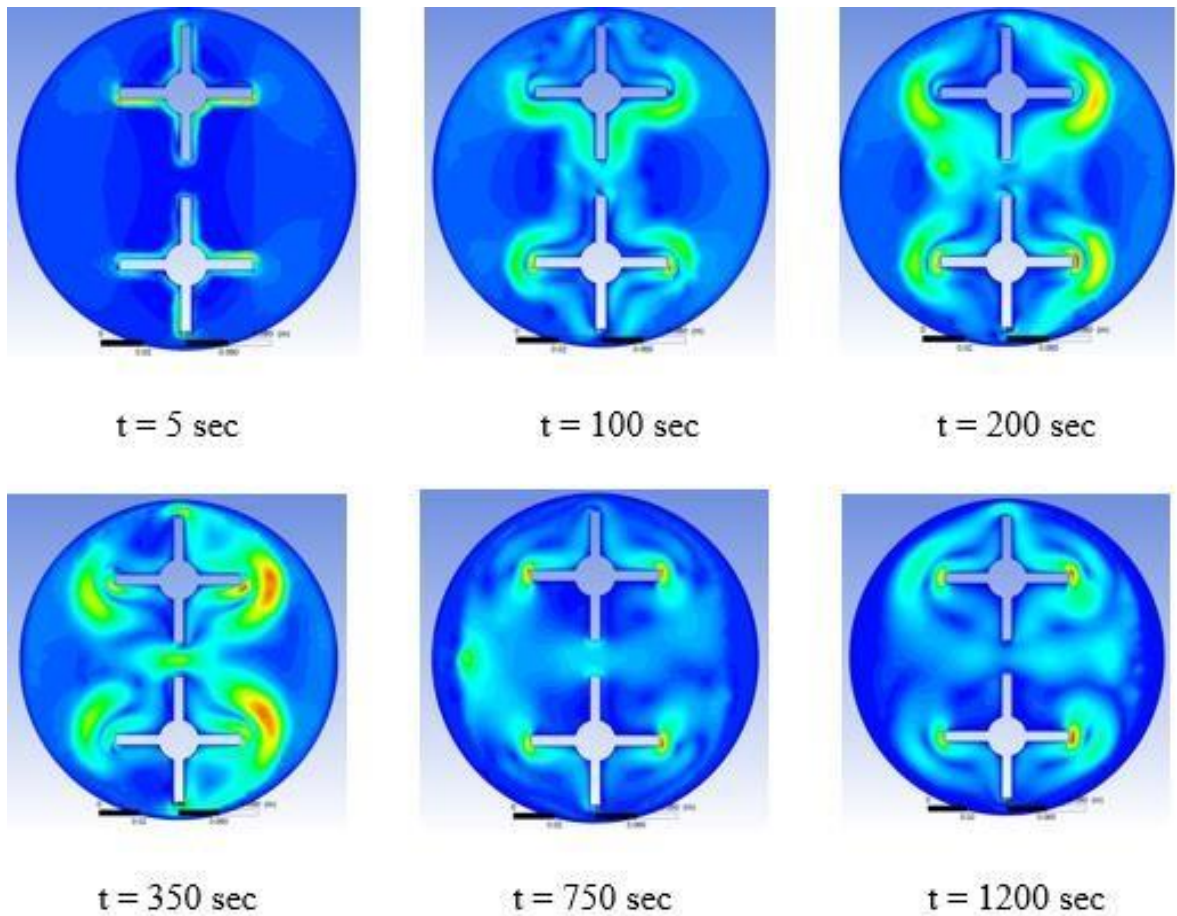

**Fig. 31.** Velocity (streamlines) contours for Model 5

# 6. Design Comparison of Models

## 6.6.1 Comparison of Liquid Fraction of Different Models

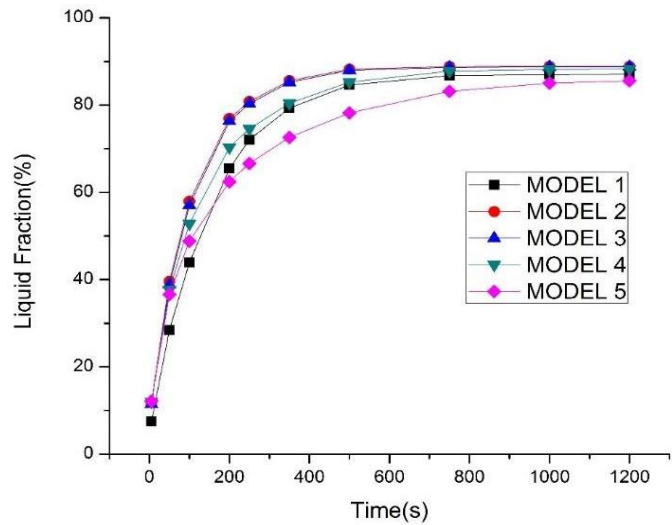

**Fig. 32.** Comparison of Liquid Fractions of Different Models

The trends of graph for all models are approximately following the same pattern.

**Table 4.** Variation of liquid fraction with time for different models

| Time (s) | Liquid Fraction (%) | | | | |
|---|---|---|---|---|---|
| | Model 1 | Model 2 | Model 3 | Model 4 | Model 5 |
| 5 | 7.411 | 11.828 | 11.433 | 11.997 | 12.071 |
| 100 | 43.887 | 57.922 | 57.01 | 52.81 | 48.837 |
| 200 | 65.45 | 76.93 | 76.296 | 70.35 | 62.4321 |
| 350 | 79.30 | 85.604 | 85.224 | 80.42 | 72.5692 |
| 750 | 86.74 | 88.837 | 88.663 | 87.78 | 83.1237 |
| 1200 | 87.038 | 88.8967 | 88.773 | 88.56 | 85.539 |

- Graph of liquid fraction clearly showing that the values of Model 1 (without fin) are lower than the values of Model 2 (with rectangular fins). This is because using the fins increases the surface area of heat transfer.
- Percentage change of Liquid fraction for 'Model 1' and 'Model 2' at 750 seconds is 2.36% and at 1200 seconds is 2.09%. The 'Model 2' is more efficient than 'Model 1'.
- In the comparison of 'Model 2' with 'Model 3', the only change is design of fins. In 'Model 2' rectangular fins used and in 'Model 3' trapezoidal fins used to keep the total surface area of heat source constant.

- Percentage change in Liquid fraction for 'Model 2' and 'Model 3' at 1200 seconds is 0.14% only. This clearly shows that the efficiency of both the models is approximately same.
- In comparison of 'Model 2' with 'Model 4', they are much differ in their design. The 'Model 4' has three heat sources with large rectangular fins size as compared to 'Model 2'.
- Percentage change of liquid fraction of 'Model 4' compared to 'Model 2' is 0.378%. This clearly shows that their efficiencies are approximately similar when operating under same conditions.
- 'Model 4' is giving the same results with three heat source as compared to 'Model 4' with four heat source. The saving in one heat source reduces the material requirement and it also reduces the cost.
- 'Model 5' consists of two heat sources and the size of rectangular fins are larger than any other design to keep the total surface area constant.
- Percentage change of liquid fraction of 'Model 2' compared to 'Model 5' at 1200 seconds is 3.78%. This clearly shows that the 'Model 2' is much efficient than 'Model 5'.
- Percentage change of Liquid fraction of 'Model 4' compared to 'Model 5' at 1200 seconds is 3.41%. This shows that the 'Model 4' is more efficient than 'Model 5'.
- The efficiency of 'Model 4' is approximately equal to 'Model 2'. This clearly shows that the 'Model 4' with three heat source is the optimized design.
- On further reducing the heat source from three to two the obtained design does not provide good results. Hence it is not optimized.

**6.6.2. Comparison of Mean Temperature of Different Models**

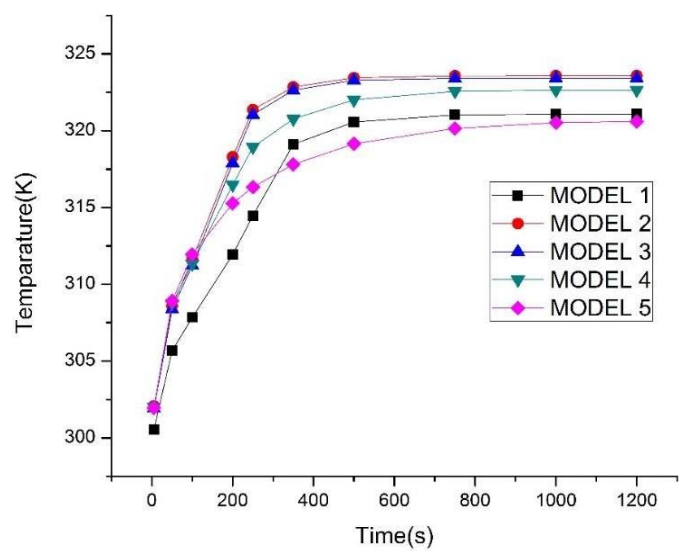

**Fig. 33.** Comparison of mean temperature of different Models

Table 5. Variation of Temperature with Time for different models

| Time (s) | Temperature (K) | | | | |
|---|---|---|---|---|---|
| | Model 1 | Model 2 | Model 3 | Model 4 | Model 5 |
| 5 | 300.5624 | 302.0769 | 301.9178 | 302.023 | 301.939 |
| 100 | 307.839 | 311.619 | 311.226 | 311.333 | 311.9458 |
| 200 | 311.944 | 318.301 | 317.8939 | 316.487 | 315.2805 |
| 350 | 319.116 | 322.847 | 322.6411 | 320.77 | 317.817 |
| 750 | 321.036 | 323.579 | 323.4128 | 322.5634 | 320.1568 |
| 1200 | 321.086 | 323.587 | 323.4214 | 322.64 | 320.6113 |

Fig. 33 shows the variation of mean temperature with time for different models.

- 'Model 5' has the lowest value of mean temperature around 320.6K at 1200 seconds.
- The values of mean temperature from start of time to end are similar in case of 'Model 2' and 'Model 3'.
- The mean temperature value of 'Model 4' at 1200 seconds is less than 'Model 2' but this design consists of only three heat source and providing the similar melting at the given time period.
- The values of mean temperature are more in case of 'Model 1' as compared to 'Model 5' this clearly shows that further decrease in number of heat source from three to two is not beneficial.
- The 'Model 4' with three heat sources provides the best optimized results because it creates the same effect of melting in given time period with lesser value of mean temperature.

## 7. Conclusion

From this investigation there are some points to show the outcome of current research that is given below:

**(a)** The operating and boundary conditions for all models were exactly similar. The temperature of heat source and fins assumed to be constant 343K for complete analysis. The outer walls of cylindrical TES assumed to be adiabatic and at atmospheric temperature of 298K.

**(b)** On comparison of percentage of liquid fraction between different models first we get 'Model 2' is more efficient than 'Model 1' due to presence of fins on surface of heat source. But on maintaining the total surface area of all designs having fins to be constant equal to 507.32 mm².

**(c)** The 'Model 3' consists of trapezoidal shaped fins with four heat sources. The 'Model 3' and 'Model 2' gives approximately same results during melting process.

**(d)** On decreasing the one heat source and maintaining the total surface area of heat source constant we get 'Model 4'. The percentage change of liquid fraction at 1200 seconds for 'Model 4' compared to 'Model 2' is 0.378%, which is very less showing approximately similar outcomes.

**(e)** Further decrease in heat source to two and increasing the size of fins by maintaining the total surface area to be constant depicted by geometry of 'Model 5'. Comparison clearly shows that 'Model 5' has 3.78% less value of liquid fraction than 'Model 2' and 3.41% less value of liquid fraction than 'Model 4'. This shows that further decrease in heat source from three to two is not efficient/beneficial.

**(f)** Results obtained by simulations clearly shows that 'Model 4' gives the best results as compared to other models and this model consists of only three heat sources saving the material required and reducing the cost also. Hence we clearly say that the 'Model 4' gives the optimized design of thermal energy storage (TES) in cylindrical form in which gallium is used as PCM.